\documentclass[english,aps, prb, twocolumn, superscriptaddress,showpacs]{revtex4-1}
\usepackage{lmodern}
\usepackage[T1]{fontenc}
\usepackage[latin9]{inputenc}
\usepackage{bm}
\usepackage{amstext}
\usepackage{graphicx}

\makeatletter

\newcommand{\lyxmathsym}[1]{\ifmmode\begingroup\def\b@ld{bold}
  \text{\ifx\math@version\b@ld\bfseries\fi#1}\endgroup\else#1\fi}

\@ifundefined{textcolor}{}
{%
 \definecolor{BLACK}{gray}{0}
 \definecolor{WHITE}{gray}{1}
 \definecolor{RED}{rgb}{1,0,0}
 \definecolor{GREEN}{rgb}{0,1,0}
 \definecolor{BLUE}{rgb}{0,0,1}
 \definecolor{CYAN}{cmyk}{1,0,0,0}
 \definecolor{MAGENTA}{cmyk}{0,1,0,0}
 \definecolor{YELLOW}{cmyk}{0,0,1,0}
 }

\makeatother

\usepackage{babel}

\begin{document}

%\title{Spin Fluctuations in Superconducting Cs$_{x}$Fe$_{2-y}$Se$_{2}$}

\title{Spin-wave excitations and superconducting resonant mode in Cs$_{x}$Fe$_{2-y}$Se$_{2}$}

\author{A. E. Taylor }
\email[]{a.taylor1@physics.ox.ac.uk}

\affiliation{Department of Physics, University of Oxford, Clarendon Laboratory,
Parks Road, Oxford, OX1 3PU, United Kingdom}

\author{R. A. Ewings}

\affiliation{ISIS Facility, Rutherford Appleton Laboratory, STFC, Chilton, Didcot,
Oxon, OX11 0QX, United Kingdom}

\author{T. G. Perring}

\affiliation{ISIS Facility, Rutherford Appleton Laboratory, STFC, Chilton, Didcot,
Oxon, OX11 0QX, United Kingdom}

\affiliation{Department of Physics and Astronomy, University College London, Gower
Street, London, WC1E 6BT, United Kingdom}

\author{J. S. White}

\affiliation{Laboratory for Neutron Scattering, Paul Scherrer Institute, CH-5232 Villigen PSI, Switzerland}

\affiliation{Laboratory for Quantum Magnetism, Ecole Polytechnique F\'{e}d\'{e}rale de Lausanne (EPFL), 1015 Lausanne, Switzerland}

\author{P. Babkevich}

\affiliation{Laboratory for Quantum Magnetism, Ecole Polytechnique F\'{e}d\'{e}rale de Lausanne (EPFL), 1015 Lausanne, Switzerland}

\author{A. Krzton-Maziopa}

\affiliation{Laboratory for Developments and Methods, Paul Scherrer Institute, CH-5232 Villigen PSI, Switzerland}

\affiliation{Faculty of Chemistry, Warsaw University of Technology, PL-00664 Warsaw, Poland}

\author{E. Pomjakushina}

\affiliation{Laboratory for Developments and Methods, Paul Scherrer Institute, CH-5232 Villigen PSI, Switzerland}

\author{K. Conder}

\affiliation{Laboratory for Developments and Methods, Paul Scherrer Institute, CH-5232 Villigen PSI, Switzerland}

%\author{B. Roessli ???}

%\affiliation{Laboratory for Neutron Scattering, Paul Scherrer Institute, CH-5232 Villigen PSI, Switzerland}

\author{A. T. Boothroyd}
\email[]{a.boothroyd@physics.ox.ac.uk}

\affiliation{Department of Physics, University of Oxford, Clarendon Laboratory,
Parks Road, Oxford, OX1 3PU, United Kingdom}
\begin{abstract}
%We report neutron inelastic scattering measurements on the normal and superconducting states of single-crystalline Cs$_{0.8}$Fe$_{1.9}$Se$_{2}$. Consistent with previous measurements on Rb$_{x}$Fe$_{2-y}$Se$_{2}$, we observe two distinct spin excitation signals: (i) spin-wave excitations characteristic of the block antiferromagnetic order found in insulating $A_{x}$Fe$_{2-y}$Se$_{2}$ compounds, and (ii) a resonance-like magnetic peak localized in energy at 11\,meV and at an in-plane wave vector of $(0.25, 0.5)$, which increases in intensity below $T_\mathrm{c} = 27\,$K. The existence of a resonance in the magnetic spectrum of Rb$_{x}$Fe$_{2-y}$Se$_{2}$ and now of Cs$_{x}$Fe$_{2-y}$Se$_{2}$ suggests that this is a common feature of superconductivity in this family. We could find no measurable response to superconductivity in the low energy spin-wave excitations in Cs$_{0.8}$Fe$_{1.9}$Se$_{2}$, consistent with the notion of spatially separate magnetic and superconducting phases.
We report neutron inelastic scattering measurements on the normal and superconducting states of single-crystalline Cs$_{0.8}$Fe$_{1.9}$Se$_{2}$. Consistent with previous measurements on Rb$_{x}$Fe$_{2-y}$Se$_{2}$, we observe two distinct spin excitation signals: (i) spin-wave excitations characteristic of the block antiferromagnetic order found in insulating $A_{x}$Fe$_{2-y}$Se$_{2}$ compounds, and (ii) a resonance-like magnetic peak localized in energy at 11\,meV and at an in-plane wave vector of $(0.25, 0.5)$. The resonance peak increases below $T_\mathrm{c} = 27\,$K, and has a similar absolute intensity to the resonance peaks observed in other Fe-based superconductors. The existence of a magnetic resonance in the spectrum of Rb$_{x}$Fe$_{2-y}$Se$_{2}$ and now of Cs$_{x}$Fe$_{2-y}$Se$_{2}$ suggests that this is a common feature of superconductivity in this family. The low energy spin-wave excitations in Cs$_{0.8}$Fe$_{1.9}$Se$_{2}$ show no measurable response to superconductivity, consistent with the notion of spatially separate magnetic and superconducting phases.
%We could find no measurable indication of coupling between the low energy spin-wave excitations and superconductivity in Cs$_{0.8}$Fe$_{1.9}$Se$_{2}$, consistent with the notion of spatially separate magnetic and superconducting phases.
%We could find no measurable (indication of) coupling between the low energy spin-wave excitations and superconductivity in Cs$_{0.8}$Fe$_{1.9}$Se$_{2}$, consistent with the notion of spatially separate magnetic and superconducting phases.
\end{abstract}

% 74.25.Ha = Magnetic properties of SCs , 74.70.Xa = (SCs) Pnictides&Chalcogenides, 78.70.Nx = Neutron inelastic scattering, 75.30.Ds = Spin Waves

\maketitle

\section{Introduction}

The $A_{x}$Fe$_{2-y}$Se$_{2}$ compounds ($A$ = K, Rb, Cs and Tl) present an interesting new
twist in the field of iron-based superconductors. The discovery of superconductivity with transition temperatures
$T{}_{\mathrm{{c}}}\approx30\, $K in this series,\cite{guo_superconductivity_2010,wang_superconductivity_2011,krzton-maziopa_synthesis_2011,Fang_TlFe2Se2_2011}
in conjunction with antiferromagnetism with an unusually high ordering temperature
$T{}_{\mathrm{{N}}}$ of up to 559$\,$K and large ordered moment of about 3.3$\,\mu_{\mathrm{B}}$
per Fe,\cite{bao_novel_2011} naturally raises the question:
can superconductivity coexist microscopically with such a robust magnetic state? Although there are regions in the phase diagrams of the iron pnictide superconductors in which magnetism and superconductivity are believed to coexist microscopically, the highest $T{}_{\mathrm{{c}}}$s and bulk superconductivity are found when the magnetic state has been suppressed.\cite{johnston_puzzle_2010,stewart_superconductivity_2011,lumsden_magnetism_2010} Another distinct feature of the $A_{x}$Fe$_{2-y}$Se$_{2}$ systems is their band structure.\cite{zhang_nodeless_2011,qian_absence_2011} The Fermi surface lacks the large hole pocket at the zone center that features prominently in theories of superconductivity and magnetism in other iron-based superconductors.

The magnetic structure observed in superconducting $A_{x}$Fe$_{2-y}$Se$_{2}$ samples consists of blocks
of four ferromagnetically aligned Fe spins, with antiferromagetic
alignment between these blocks. This magnetic state forms on a $\sqrt{5}\times\sqrt{5}$
superstructure of ordered Fe vacancies
that has optimal composition $A_{0.8}$Fe$_{1.6}$Se$_{2}$.\cite{pomjakushin_iron-vacancy_2011,bao_novel_2011,pomjakushin_room_2011,wang_antiferromagnetic_2011,ye_common_2011} A $\sqrt{2}\times\sqrt{2}$
ordered phase has also been observed in some samples and is thought to be closely related to the superconducting phase.\cite{wang_antiferromagnetic_2011,ricci_intrinsic_2011,li_kfe_2se_2_2012}

%Initial investigations of $A_{x}$Fe$_{2-y}$Se$_{2}$ supported a picture of microscopic coexistence of the superconducting and antiferromagnetic states. As well as estimates of 25--100\% superconducting volume fraction \cite{guo_superconductivity_2010,krzton-maziopa_synthesis_2011}, muon spin rotation ($\mu$-SR)\cite{shermadini_coexistence_2011}, nuclear magnetic resonance (NMR)\cite{torchetti_^77se_2011,yu_^77se_2011,ma_superconducting_2011,kotegawa_possible_2011} and neutron diffraction \cite{bao_novel_2011} studies all found evidence for coexistence and were backed up by calculations based on the antiferromagnetic state \cite{das_modulated_2011,yan_electronic_2011,cao_electronic_2011,zhang_superconductivity_2011}. Further work, however, has found evidence for a spatial separation of superconducting (metallic) and antiferromagnetic (insulating or semiconducting) phases from scanning tunnelling microscopy (STM)\cite{li_phase_2011}, optical\cite{charnukha_optical_2012}, scanning nanofocus x-ray diffraction\cite{ricci_nanoscale_2011}, transmission electron microscopy\cite{chen_electronic_2011}, neutron diffraction\cite{wang_antiferromagnetic_2011} and M�ssbauer\cite{ksenofontov_phase_2011} experiments.

Initial experimental investigations of $A_{x}$Fe$_{2-y}$Se$_{2}$ supported a picture of microscopic coexistence
of the superconducting and antiferromagnetic states.\cite{guo_superconductivity_2010,krzton-maziopa_synthesis_2011,shermadini_coexistence_2011,torchetti_^77se_2011,yu_^77se_2011,ma_superconducting_2011,kotegawa_possible_2011,bao_novel_2011}
These studies were backed up by calculations based on the antiferromagnetic
state.\cite{das_modulated_2011,yan_electronic_2011,cao_electronic_2011,zhang_superconductivity_2011}
Further work, however, has found evidence for a spatial separation of superconducting (metallic)
and antiferromagnetic (insulating or semiconducting) phases.\cite{li_phase_2011,charnukha_optical_2012,ricci_nanoscale_2011,chen_electronic_2011,wang_antiferromagnetic_2011,ksenofontov_phase_2011} The most recent results, from NMR,\cite{texier_nmr_2012} scanning electron microscopy (SEM),\cite{speller_microstructural_2012}
optical spectroscopy,\cite{yuan_nanoscale_2012} Raman scattering and optical microscopy,\cite{zhang_two-magnon_2012} and low energy muon spin rotation\cite{charnukha_PRL_2012} may help to explain the apparent discrepancies in the earlier work. They indicate that phase separation occurs with a complex plate-like morphology on a sub-micron scale. Proximity effects between nanodomains could therefore allow an interplay between superconducting and magnetic regions, and may explain
the apparent bulk superconductivity despite estimates of a genuine
superconducting phase fraction of only 5--10\%.

The interplay between superconductivity and static magnetic order remains a key issue in the $A_{x}$Fe$_{2-y}$Se$_{2}$ family. Another important property is the magnetic dynamics, which are widely thought to play a role in mediating superconductivity in the iron-based superconductors.\cite{johnston_puzzle_2010,stewart_superconductivity_2011,lumsden_magnetism_2010} Up to now, investigations of the magnetic dynamics have focussed
on Rb$_{x}$Fe$_{2-y}$Se$_{2}$. The spin-wave spectrum of the insulating parent antiferromagnetic phase has been measured by inelastic neutron scattering
and the results were successfully modelled in terms of a local moment Heisenberg Hamiltonian.\cite{wang_spin_2011} Superconducting samples of Rb$_{x}$Fe$_{2-y}$Se$_{2}$ have also been studied, and a spin resonance has been discovered.\cite{park_magnetic_2011} The resonance is quasi-two-dimensional and characterized by an increase in scattering intensity below $T_{\rm c}$ at an energy of approximately 14\,meV and at the wave vector ${\bf Q}=(0.25, 0.5)$ and equivalent positions, which corresponds to $(\pi/2, \pi)$ in square lattice notation.\cite{park_symmetry_2010} This wave vector is not the same as the usual resonance wave vector of the iron based superconductors, which is ${\bf Q}=(0.5,0)$ etc. In Ref.~\onlinecite{park_magnetic_2011} it was suggested that the position of the resonance in Rb$_{x}$Fe$_{2-y}$Se$_{2}$ can be traced to the nesting of electron-like Fermi surface pockets together with a $d$-wave superconducting pairing state based on the theory of Maier \emph{et al.},\cite{maier_d-wave_2011} unlike the $s_{\pm}$ pairing generally thought to be present in other iron-based superconductors. In another study of Rb$_{x}$Fe$_{2-y}$Se$_{2}$, a magnetic signal was reported close to ${\bf Q}=(0.5,0)$ in addition to spin-wave excitations from the block antiferromagnetic order and the magnetic resonance at ${\bf Q}=(0.5,0.25)$.\cite{wang_pi0_2012}

In this work we studied the spin excitations in superconducting Cs$_{x}$Fe$_{2-y}$Se$_{2}$, with particular focus on the low energy magnetic features and their response to superconductivity. We find that the spin excitations associated with the block antiferromagnetic order have a very similar spectrum to those observed in non-superconducting Rb$_{x}$Fe$_{2-y}$Se$_{2}$ up to the highest energy probed ($\sim$ 150\,meV).  To within experimental error, we find no influence of superconductivity on the low energy magnetic excitations from the block antiferromagnetic order, in contrast to the response of the magnetic Bragg peak and a two-magnon Raman peak which both show a small anomaly in intensity on cooling below $T_{\rm c}$.\cite{bao_novel_2011,zhang_two-magnon_2012}  Finally, we observe a spin resonance below $T_{\rm c}$ at $(0.25,0.5)$, establishing that this feature is not confined to Rb$_{x}$Fe$_{2-y}$Se$_{2}$ but is present in other members of the $A_{x}$Fe$_{2-y}$Se$_{2}$ family. The results suggest that the sample consists of distinct magnetically ordered and superconducting phases.

%In this work we probe the interplay between superconductivity and magnetism via the magnetic dynamics of each phase. We do this by looking at the temperature dependence of the magnetic fluctuations that can be ascribed to each phase. Previously the temperature dependence of the antiferromagnetic phase was probed only via the static magnetism - i.e. the Bragg signal was measured and its intensity found to saturate at $T{}_{\mathrm{{c}}}$\cite{bao_novel_2011}. We measure the spin fluctuations in Cs$_{x}$Fe$_{2-y}$Se$_{2}$ finding 1) a spin wave spectra associated purely with the magnetic phase with no significant change in intensity at $T{}_{\mathrm{{c}}}$ and similar to that seen in insulating Rb$_{x}$Fe$_{2-y}$Se$_{2}$ and 2) the spin resonance at $\bm{{Q}}=(0.5,\,0.25)$ seemingly linked just to the superconducting phase which increases in intensity at $T{}_{\mathrm{{c}}}$. This provides evidence that the spin interactions across the range of A$_{x}$Fe$_{2-y}$Se$_{2}$ systems are similar, and that the $\bm{{Q}}=(0.5,\,0.25)$ fluctuations exist in the superconducting phase and are key to the superconducting mechanism.

\section{Experimental Methods}

\begin{figure}
\includegraphics[clip,width=0.8\columnwidth]{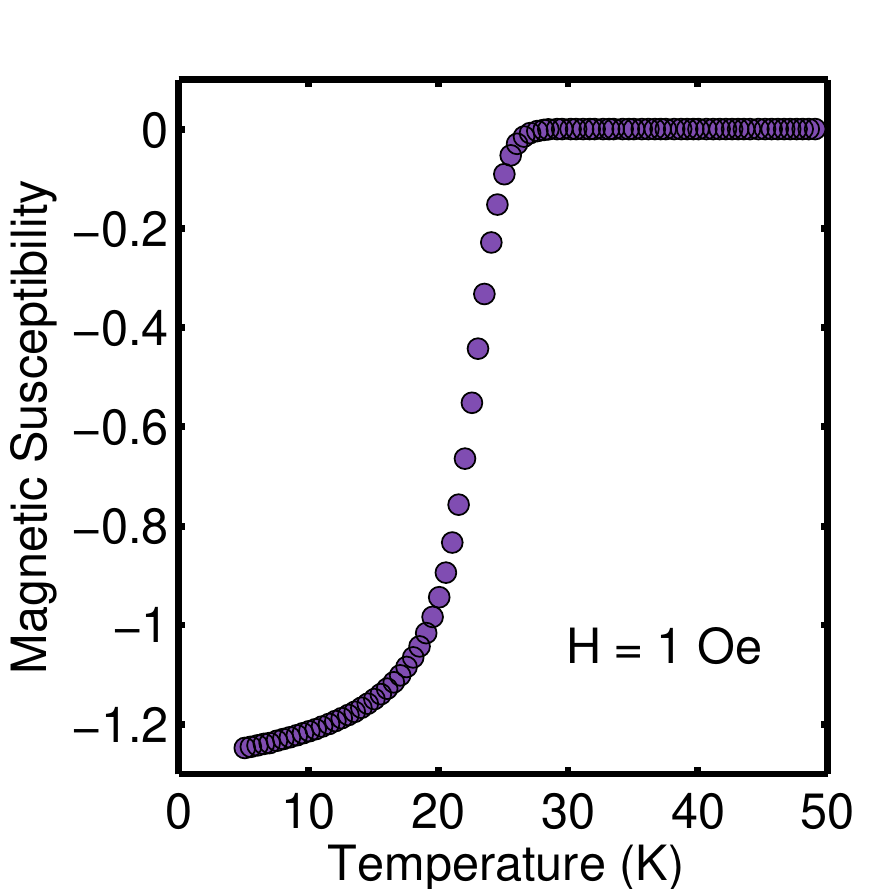}

\caption{\label{fig:Squid}(Color online) Magnetic susceptibility of one of the Cs$_{0.8}$Fe$_{1.9}$Se$_{2}$
single crystals used here, measured with a field of $1\,$Oe applied along the $c$ axis after cooling in zero field. A susceptibility of $-1$ corresponds to full Meissner flux exclusion, but as no demagnetization corrections have been applied $-1$ is not a rigorous lower bound.}

\end{figure}

The Cs$_{x}$Fe$_{2-y}$Se$_{2}$ single crystals were grown by the
Bridgman process as decribed in Ref.~\onlinecite{krzton-maziopa_synthesis_2011}.
The nominal composition of the crystals used in this study is Cs$_{0.8}$Fe$_{1.9}$Se$_{2}$,
and their superconducting and magnetic properties have been reported
previously\cite{krzton-maziopa_synthesis_2011,shermadini_coexistence_2011,pomjakushin_iron-vacancy_2011}.
The crystals were coated in Cytop varnish
before handling in air, and then checked for crystalline quality prior to the experiment.
Magnetic susceptibility measurements shown in Fig.~\ref{fig:Squid} established that the onset of bulk superconductivity occurs at $T{}_{\mathrm{{c}}}=27\,$K. By `bulk' we mean that full flux exclusion is achieved after cooling in zero field. However, this does not necessarily imply 100\% superconducting volume fraction, since non-superconducting regions can be screened by surface currents in a zero-field-cooled measurement. A crystal from the neutron scattering sample was remeasured after the experiment and found to have an unchanged $T_{\rm c}$.

The inelastic neutron scattering experiments were performed on the MERLIN time-of-flight (TOF) chopper spectrometer at the ISIS Facility.\cite{bewley_merlin_2006} Three single crystals were co-aligned to give a sample of total mass 0.42\,g, with a uniform mosaic of $2.5^{\circ}$ (full width at half maximum). The sample was mounted with the $c$ axis parallel to the incident neutron beam, and the $a$ axis horizontal. Spectra were recorded in the large position-sensitive detector array with neutrons of incident energy $E_{\rm i}=33, 40, 50, 60, 100$ and $180$\,meV at $T=4$\,K, and $E_{\rm i}=33$\,meV at $T=4$, 20, 34 and 44\,K. For a fixed sample orientation only three of the four $({\bf Q},E)$ components are independent. We will use $E$ and the two in-plane wave vector components $(H,K)$. This means that the out-of-plane wave vector component varies with $E$.  The scattering from a standard vanadium sample was used to normalize the spectra and place them on an absolute intensity scale, with units mb\,sr$^{-1}$\,meV$^{-1}$\,f.u.$^{-1}$, where 1\,mb = 10$^{-31}$\,m$^{2}$ and f.u. stands for formula unit of Cs$_{0.8}$Fe$_{1.9}$Se$_{2}$.

\section{Results}

\begin{figure*}
\includegraphics[clip,width=0.9\textwidth]{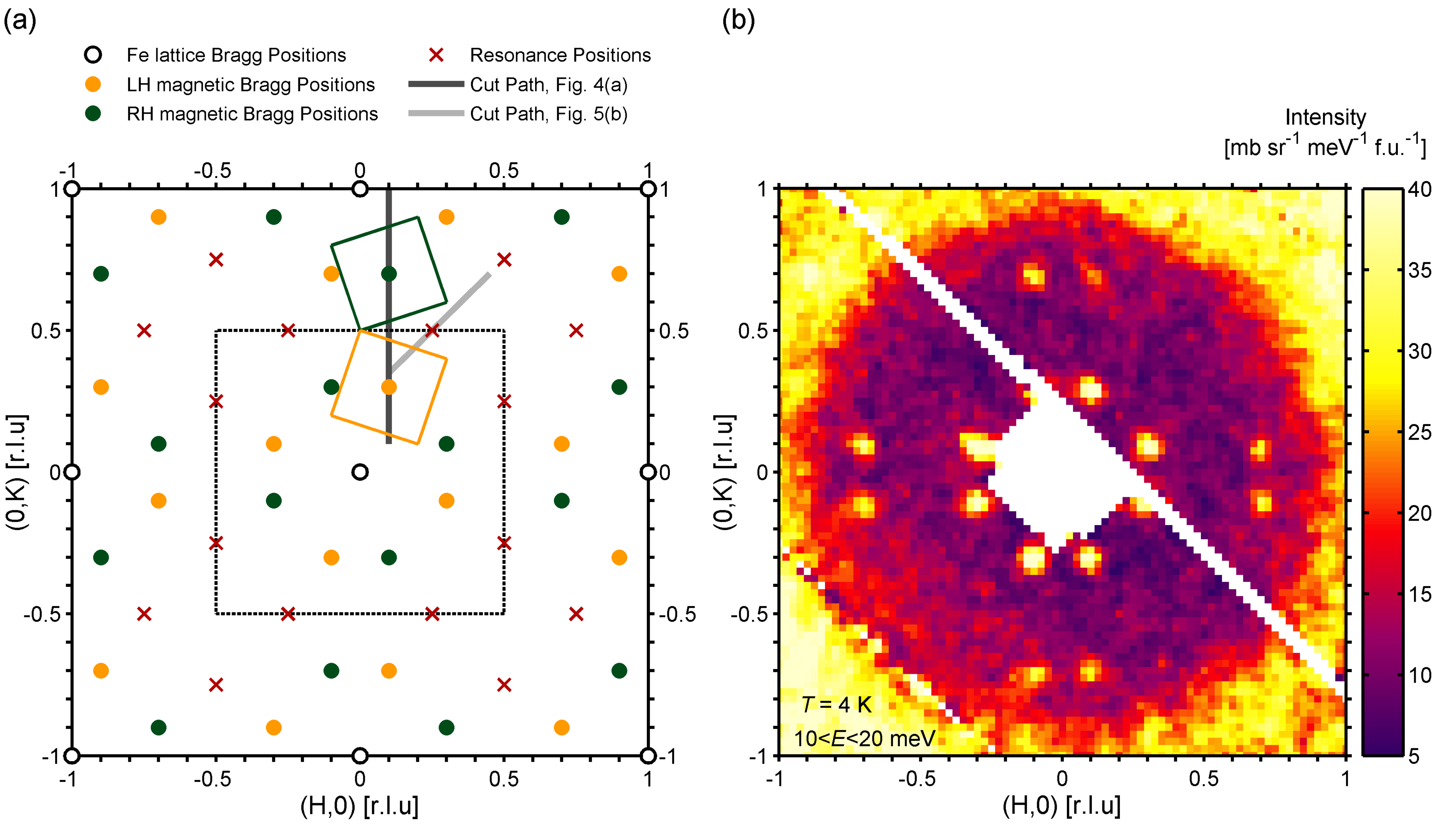}

\caption{\label{fig:Map}(Color online) (a) Map of two-dimensional reciprocal space for Cs$_{x}$Fe$_{2-y}$Se$_{2}$, showing positions of the magnetic ordering vectors and the magnetic resonance signals. Wave vectors are indexed with respect to the one-Fe unit cell (unfolded Brillouin zone). The open circles are reciprocal lattice points, and the dashed square is the first Brillouin zone for this unit cell. The two small tilted squares are the first Brillouin zones for the unit cells of the left-handed (LH) and right-handed (RH) magnetic domains.\cite{bao_novel_2011} The paths along which cuts shown in later figures were taken are represented by thick solid lines. (b) Neutron scattering intensity map of Cs$_{0.8}$Fe$_{1.9}$Se$_{2}$ in the same area of reciprocal space as shown in (a). The data were recorded with an incident neutron energy of 100\,meV. The areas of missing data are due to the beam stop and masked detectors.}

\end{figure*}

Figure~\ref{fig:Map}(a) is a map of the $(H,K)$ plane in two-dimensional reciprocal space, showing the positions of the antiferromagnetic Bragg peaks and the magnetic resonance signal reported in Ref.~\onlinecite{park_magnetic_2011}. We index positions in reciprocal space with respect to the one-Fe sub-lattice, lattice parameters $a = b = 2.8$\,{\AA}. Figure 2(b) is a map of the neutron scattering intensity averaged over the energy range 10 to 20\,meV and projected onto the same region of the $(H,K)$ plane as shown in Fig.~\ref{fig:Map}(a). The strong scattering signal localized at $\mathbf{Q}_{\rm AFM}=(0.1,0.3)$ and equivalent positions is due to magnetic fluctuations associated with the block antiferromagnetic order on the $\sqrt{5}\times\sqrt{5}$ Fe vacancy superstructure. The eight-fold symmetry of the magnetic spectrum, which derives from the superposition of two four-fold patterns from left-handed and right-handed magnetic structures, respectively, is apparent from this figure. All spectra presented hereafter have been folded into one octant to improve statistics.

%Figure \ref{fig:Map} (b) shows a slice through the data at low energies. A map of reciprocal space in the Fe1 sub lattice cell notation is presented in Fig. \ref{fig:Map} (a) that can be directly compared to the data. The strong scattering signal at positions $\mathbf{Q}_{AFM}=(0.1\,0.3)$ and equivalent is due to magnetic fluctuations associated with the $\sqrt{5}\times\sqrt{5}$ superstructure. The 8-fold symmetry of the data is apparent from this figure and all subsequent data presented has been symmetrised into one octant to improve statistics. The magnetic spectrum is similar to that of insulating Rb$_{x}$Fe$_{2-y}$Se$_{2}$\cite{wang_spin_2011} with intensity dispersing out of the $\mathbf{Q}_{AFM}$ positions.

\begin{figure}
\includegraphics[clip,width=0.8\columnwidth]{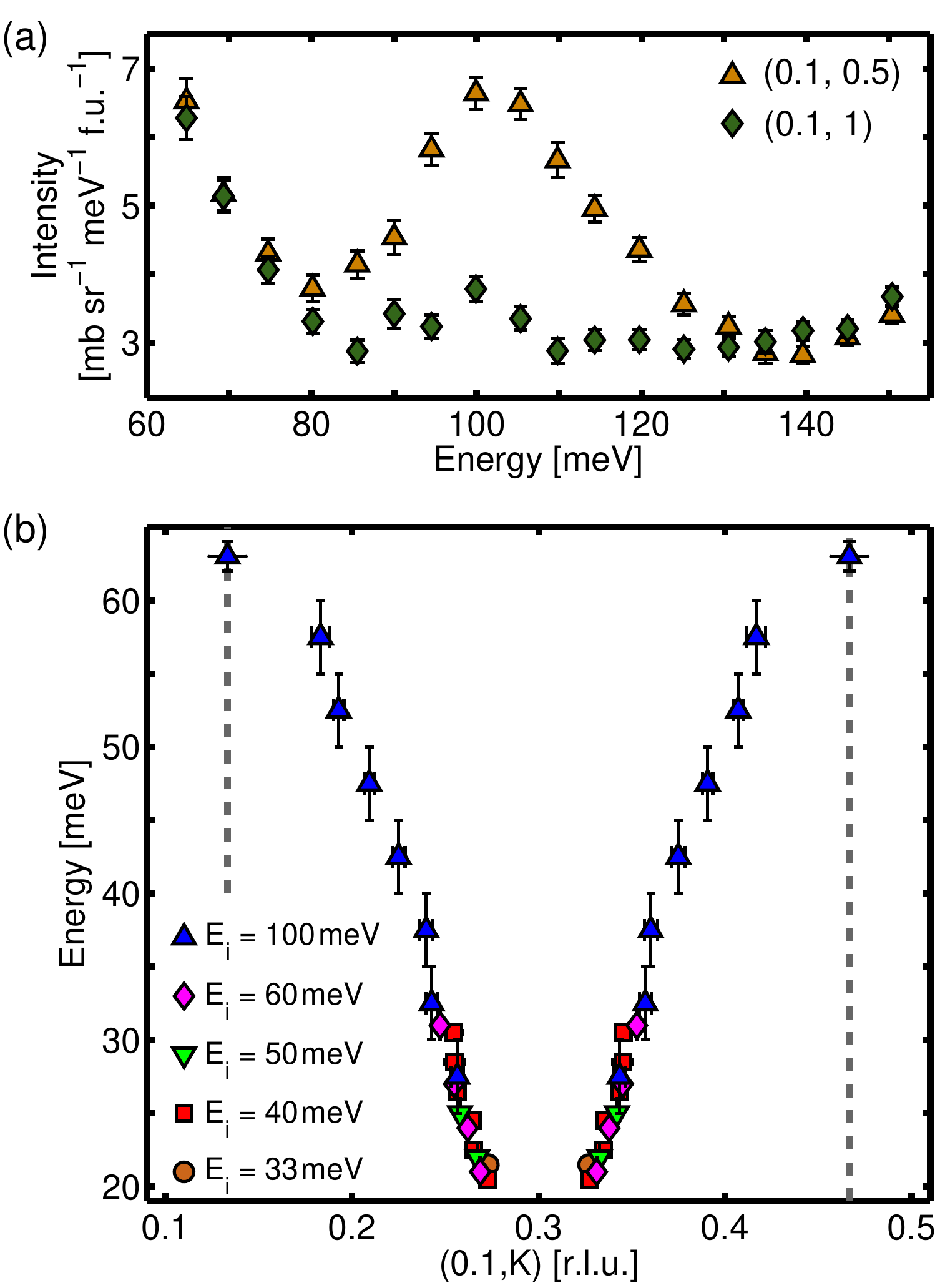}

\caption{\label{fig:Dispersion}(Color online) Spin-wave spectrum of Cs$_{x}$Fe$_{2-y}$Se$_{2}$. (a) Energy cuts showing the band of magnetic scattering around 100\,meV. The data are from a run with incident neutron energy $E_{\rm i} =180$\,meV at 4\,K. The orange triangles were recorded at the wave vector $(0.1, 0.5)$, where there is a clear magnetic signal with maximum intensity near 105\,meV. The green diamonds are a similar cut from the nearby position $(0.1, 1)$, which is away from any magnetic scattering. (b) Spin-wave dispersion of the low-energy band, measured at $4\,$K. The method used to obtain the data points is described in the text. The different colored symbols indicate data obtained with different incident neutron energies, $E_{\rm i}$. The dashed lines mark the magnetic Brillouin zone boundaries. For energies below 60\,meV, the vertical error bars represent the width of the cut in energy and the horizontal error bars represent the error in the fitted peak position. For the two points at 63\,meV, the horizontal error bar is the width of the cut in wave vector and the vertical error bar is the error from the fit.}

\end{figure}

The magnetic spectrum is revealed in more detail in Fig.~\ref{fig:Dispersion}, which shows a strongly dispersive spin-wave band extending from below 20\,meV up to 63\,meV, and a second band between 85 and 120\,meV. The existence of the latter is demonstrated in Fig.~\ref{fig:Dispersion}(a) via two energy scans recorded at fixed wavevectors of $(0.1, 0.5)$ and $(0.1, 1)$. These positions were chosen after inspection of an intensity map like that in Fig.~\ref{fig:Map}(b) but at an energy of 100\,meV, which showed a regular pattern of diffuse magnetic scattering with maximum intensity at $(0.1, 0.5)$ and minimum at $(0.1, 1)$.

Figure~\ref{fig:Dispersion}(b) plots the in-plane dispersion of the lower spin-wave band. In constant-energy maps in the $(H,K)$ plane, the low-energy spin-wave scattering appears as a ring of intensity centered on the ${\bf Q}_{\rm AFM}$ positions. The points in Fig.~\ref{fig:Dispersion}(b) were obtained as follows. Gaussian fits were made to peaks in constant-energy cuts along the line $(0.1,K)$ passing through the magnetic wavevectors ${\bf Q}_{\rm AFM} = (0.1, 0.3)$ and $(0.1, 0.7)$ --- see Fig.~\ref{fig:Map}(a). Gaussian functions fitted to pairs of peaks symmetrically displaced either side of each ${\bf Q}_{\rm AFM}$ were constrained to have the same area and width. The peak positions were corrected for the systematic shift caused by the curvature of the dispersion surface over the width $\Delta H$ of the cuts. Where appropriate, a non-magnetic background was estimated from cuts taken along nearby lines in reciprocal space. The points at the magnetic Brillioun zone (BZ) boundaries (marked in Fig.~\ref{fig:Dispersion}(b) by dashed lines) were obtained from a Gaussian fit to the peak in a background-corrected energy cut.

%The magnetic spectrum is revealed in more detail in Fig.~\ref{fig:Dispersion}, where the dispersion is plotted. Apart from the final point at the Brillioun Zone boundary, the points in Fig. \ref{fig:Dispersion} (b) were found by taking constant energy cuts through the data at position $(0.1\,\mathrm{{K})}$ and where appropriate subtracting the background from a nearby point in reciprocal space with no magnetic scattering. The vertical errorbars represent the width of the cut in energy, and the horizontal errorbars represent the error from the fitting. Gaussians were fit to the peaks along the K direction to find the radius of the ring dispersing out from the Bragg positions $(0.1\,0.3)$ and $(0.1\,0.7)$ (see Fig. \ref{fig:Map} (a)). The radius found was corrected for the effect of the width of the cut in H. The Gaussians symmetrically on either side of each Bragg position were contrained to have the same area and sigma. An overall flat background was also allowed to vary and found to be zero where a non-magnetic background subtraction had been applied. The points at the magnetic Brillioun zone (BZ) boundary (marked in the Fig. \ref{fig:Dispersion} (b) by the dashed lines) were found by doing a cut along energy at the magnetic BZ edge position, and subtracting a similar cut from a nearby non-magnetic position. A peak in energy was found and fitted with a Gaussian function.

\begin{figure}
\includegraphics[clip,width=0.9\columnwidth]{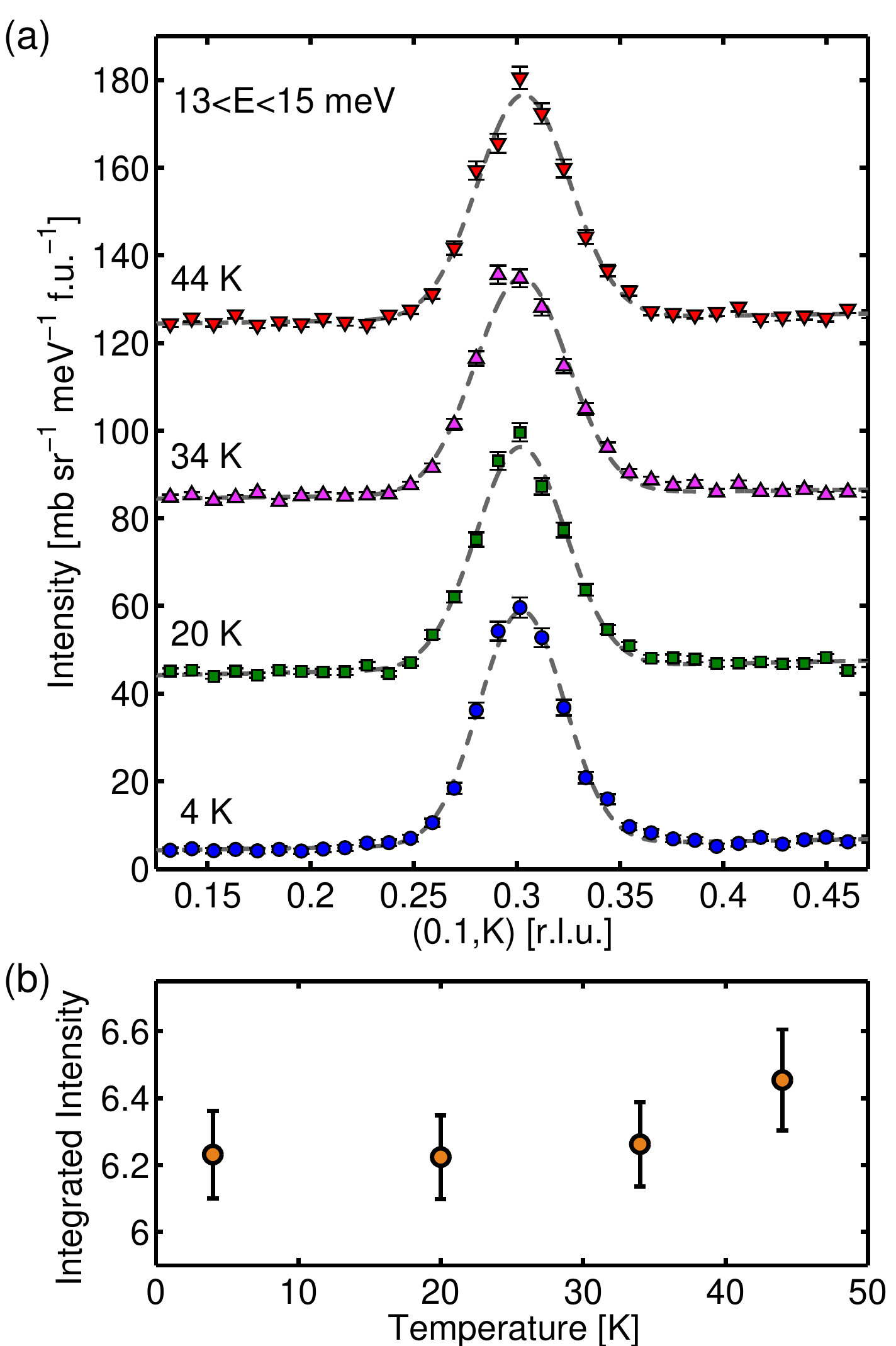}

\caption{\label{fig:AFM_peaks}(Color online) (a) Constant-energy cuts through ${\bf Q}_{\rm AFM}=(0.1,0.3)$ showing the temperature dependence of the spin-wave scattering averaged over the energy range 13 to 15\,meV. Data were recorded with an incident energy $E_{\rm i}=33$\,meV, giving an out-of-plane wave vector component $L=1.28$. Successive cuts are displaced vertically by 40 units for clarity. Dashed lines are fits to Gaussian peaks on a linear background, as described in the main text. (b) The integrated intensity (in mb$\,$sr$^{-1}$meV$^{-1}$f.u.$^{-1}$$\mathrm{{\AA}^{-1}}$) given by the area of the fitted Gaussian peaks as a function of temperature. The intensities have been normalized by the Bose factor $[1-\exp(-E/k_{\rm B}T)]^{-1}$.}

\end{figure}

The results shown in Figs.~\ref{fig:Dispersion}(a) and (b) bear a very close resemblance to the magnetic spectrum of non-superconducting Rb$_{x}$Fe$_{2-y}$Se$_{2}$ reported in Ref.~\onlinecite{wang_spin_2011}. Our data are not sufficient to determine the detailed dispersion in the out-of-plane direction $(0,0,L)$, but the spectra measured with different $E_{\rm i}$ to probe ${\bf Q}_{\rm AFM}$ at different $L$ values are consistent with a minimum anisotropy gap of $7\pm1\,$meV and a maximum of about 20\,meV, somewhat lower than the maximum of 30\,meV reported\cite{wang_spin_2011} for Rb$_{x}$Fe$_{2-y}$Se$_{2}$.  In Ref.~\onlinecite{wang_spin_2011}, a third spin-wave band was observed in Rb$_{x}$Fe$_{2-y}$Se$_{2}$, with a dispersion from 180 and 230\,meV. Our data do not extend high enough in energy to confirm the existence of this band in Cs$_{x}$Fe$_{2-y}$Se$_{2}$.

%As is seen from Fig. \ref{fig:Dispersion} the low energy band of the system disperses out from magnetic zone centres $\mathbf{Q}_{AFM}$ and reaches the Brillioun zone boundary at $\sim63\,$meV. There is a spin gap $\sim63$ to $\sim100\,$meV and the next energy band appears to be relatively flat and centred on $105\,$meV, as shown in \ref{fig:Dispersion} (a). The dispersion of the lower band and position of the second band bear a very close resemblance to what is seen in Rb$_{x}$Fe$_{2-y}$Se$_{2}$\cite{wang_spin_2011}, we do not have high incident neutron energy data to confirm the existence of a third energy band at $\sim200\,$meV.

In Fig.~\ref{fig:AFM_peaks} we show the temperature dependence of the spin-wave peak at ${\bf Q}_{\rm AFM} = (0.1, 0.3)$ averaged over the energy range 13 to 15\,meV.  Figure~\ref{fig:AFM_peaks}(a) shows wave vector scans recorded at four different temperatures, two below $T_{\rm c}$ and two above $T_{\rm c}$. The peaks show no discernible change within this temperature range. To check this quantitatively we fitted the data to a Gaussian function on a linear background, allowing the width, center and area of the Gaussian, and the slope and  intercept of the background to vary. To correct for the increase in signal due to the thermal population of spin-waves we normalized the data by the Bose population factor. Figure~\ref{fig:AFM_peaks}(b) plots the areas of the fitted peaks as a function of temperature. To within the experimental error (about 3\%) there is no change upon crossing the superconducting transition temperature.

%The temperature dependence of the low energy spin dynamics from the $\mathbf{Q}_{AFM}$ points is shown in \prettyref{fig:AFM_peaks} which presents wave vector scans at four different temperatures. These wave vector scans have been corrected for a Bose population factor, and are shifted vertically in the plot for clarity. To analyze the peak quantitatively we fitted the cuts to a Gaussian function on a flat background. Fits were made in which the width, center, and amplitude of the Gaussian and the flat background were allowed to vary. The area of the Gaussians gives the integrated intensities of the peaks which represents the strength of the magnetic signal. No appreciable change in the intergrated intensity is seen upon crossing the superconducting transition temperature.

%

\begin{figure*}
\includegraphics[clip,width=1.9\columnwidth]{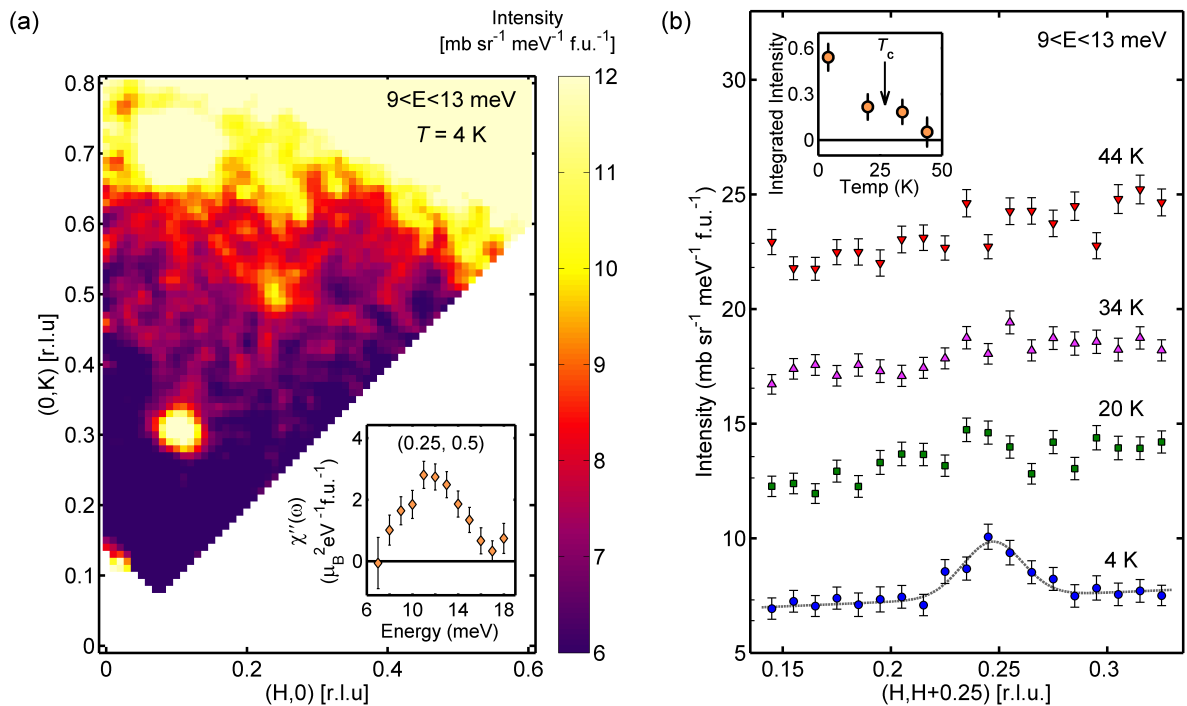}

\caption{\label{fig:ResonancePlot}(Color online) (a) Neutron scattering intensity map of Cs$_{x}$Fe$_{2-y}$Se$_{2}$ averaged over energies between 9 and 13\,meV. The data were recorded at $4\,$K with an incident energy $E_{\rm i}=33$\,meV. The inset shows the peak intensity as a function of energy, where the intensity is given as the imaginary part of the local susceptibility. (b) Constant-energy cuts through the map in (a) showing the temperature dependence of the magnetic signal at $(0.25,0.5)$. The out-of-plane wave vector at the peak is $L = 2.4$. Successive cuts are displaced vertically by 5 units for clarity. The dotted line through the 4\,K cut is a fit to a Gaussian peak on a linear background. The inset in (b) shows the integrated intensity of the 4\,K Gaussian peak and of constrained Gaussian fits to the higher temperature data (see text) as a function of temperature. The integrated intensities (in mb\,sr$^{-1}\,$meV$^{-1}\,$f.u.$^{-1}$\,\AA$^{-1}$) have been normalized by the Bose factor $[1-\exp(-E/k_{\rm B}T)]^{-1}$.}
\end{figure*}

Finally, we consider the magnetic dynamics at wave vectors away from the ${\bf Q}_{\rm AFM}$ points in reciprocal space. Figure~\ref{fig:ResonancePlot}(a) shows an intensity map recorded at 4\,K and averaged over the energy range 9 to 13\,meV. The data have been folded onto an octant of reciprocal space to improve statistics. To within the experimental error, there is no evidence for the excitations observed near $(0,0.5)$ and equivalent positions by Wang {\it et al.} in similar measurements on superconducting Rb$_{x}$Fe$_{2-y}$Se$_{2}$.\cite{wang_pi0_2012} However, our data do reveal a weak signal centered on $(0.25, 0.5)$ with a maximum at an energy of about 11\,meV (see inset to Fig.~\ref{fig:ResonancePlot}(a)).  Wave vector cuts through this peak in the $(H,H)$ direction averaged over 9--13\,meV are shown in Fig.~\ref{fig:ResonancePlot}(b) at a series of temperatures. The cut at 4\,K, well below $T_{\rm c}$, shows a well defined peak which has been fitted with a Gaussian function on a linear background (dashed line).  Above $T_{\rm c}$ the peak is either strongly suppressed or absent. Fits were made to the cuts at higher temperatures with the width and center of the Gaussian fixed to the values found at $4\,$K. The inset to Fig.~\ref{fig:ResonancePlot}(b) shows the integrated intensity of the fitted Gaussian peaks as a function of temperature. The signal clearly increases as the temperature decreases. To determine the absolute strength of the peak we have converted its integrated intensity into the $\bf Q$-averaged or local susceptibility $\chi''(\omega)$.\cite{lester_dispersive_2010} We assumed the peak is two-dimensional and used the dipole form factor of Fe$^{2+}$. The inset to Fig.~\ref{fig:ResonancePlot}(a) shows the energy dependence of $\chi''(\omega)$ at $T = 4$\,K.

\section{Discussion}

One of the goals of this work was to determine whether the spin dynamics of the block antiferromagnetic phase in superconducting samples of $A_{x}$Fe$_{2-y}$Se$_{2}$ are different to those in insulating samples, and whether they respond to superconductivity. Figure~\ref{fig:Dispersion} presents a clear demonstration that the antiferromagnetic spin-waves persist in superconducting Cs$_{x}$Fe$_{2-y}$Se$_{2}$ and have a similar spectrum to that of insulating Rb$_{x}$Fe$_{2-y}$Se$_{2}$.\cite{wang_spin_2011} We find the top of the low energy acoustic spin-wave branch to be $63\pm1$\,meV, and the center of the medium energy band to be $105\pm5$\,meV, compared with $\sim67$\,meV and $\sim115$\,meV, respectively, found in Rb$_{x}$Fe$_{2-y}$Se$_{2}$.\cite{wang_spin_2011}

We find no evidence for a coupling between the low energy spin-waves and superconductivity. This is illustrated in Fig.~\ref{fig:AFM_peaks} for an energy near 14\,meV where the scattering is strongest. However, we also examined the data from 8\,meV up to 27\,meV and found no change in the spin-wave scattering on cooling through $T_{\rm c}$ at any energy in this range.
%The temperature dependence of the spin-wave signal near 14\,meV, where the scattering is strongest, is displayed in Fig.~\ref{fig:AFM_peaks}, but we also examined a wider energy range from 8\,meV up to 27\,meV and found no change in the spin-wave scattering on cooling through $T_{\rm c}$.
From our results we can rule out any superconductivity-induced change at low energies greater than 3--4\%. By contrast, previous studies on superconducting K$_{x}$Fe$_{2-y}$Se$_{2}$ reported systematic reductions of 5\% or more in the intensities of a magnetic Bragg peak and a two-magnon Raman peak at $\sim 200$\,meV on cooling below $T_{\rm c}$.\cite{bao_novel_2011,zhang_two-magnon_2012} One possibility is that the size of the effect depends on the energy probed, however a more plausible explanation is based on the notion that these samples are phase-separated on a nanoscale into superconducting and magnetically ordered (non-superconducting) regions which only interact at the interfaces.\cite{yuan_nanoscale_2012,texier_nmr_2012,speller_microstructural_2012} Below $T_{\rm c}$, the superconducting proximity effect could suppress magnetic order near the phase boundaries, so that samples with different interfacial surface areas would respond to superconductivity by different amounts.

Although we find no effect of superconductivity on the magnetic excitations associated with the block antiferromagnetic order, we do observe the magnetic resonance peak at $(0.25,0.5)$ previously reported in superconducting Rb$_{x}$Fe$_{2-y}$Se$_{2}$.\cite{park_magnetic_2011,friemel_reciprocal-space_2012,wang_pi0_2012} As shown in Fig.~\ref{fig:ResonancePlot}, we find that the magnetic signal at $(0.25,0.5)$ increases in intensity on cooling below $T_{\rm c}$, and there is tentative evidence that the peak persists at temperatures above $T_{\rm c}$ in agreement with the observations of Friemel \emph{et al.}\cite{friemel_reciprocal-space_2012} The existence of resonance peaks in the iron pnictides has been explained in terms of nesting features in the Fermi surface enhanced by electronic correlations and superconducting coherence effects.\cite{maier_theory_2008} Within this framework, and with a realistic band structure model, Friemel \emph{et al.}\cite{friemel_reciprocal-space_2012} were able to reproduce the position of the magnetic resonance in Rb$_{x}$Fe$_{2-y}$Se$_{2}$ assuming a $d_{x^2-y^2}$ superconducting gap. Further theoretical work is needed to understand the magnetic resonance in detail, but our results at least establish that the $(0.25,0.5)$ resonance is present in another $A_{x}$Fe$_{2-y}$Se$_{2}$ superconductor. This suggests that the resonance could be a characteristic feature of superconductivity in this family.

%The measurement of the superconducting phase magnetic dynamics is then the focus of Figure \ref{fig:ResonancePlot}. In Fig. \ref{fig:ResonancePlot} (a) the resonance signal is clear as a localised peak in scattering intensity about the $(0.25\,0.5)$ point, similar to that found in superconducting Rb$_{x}$Fe$_{2-y}$Se$_{2}$\cite{park_magnetic_2011}. We find that in this case the spin dynamics are heavily dependent on the temperature, see Fig. \ref{fig:ResonancePlot} (b) and inset of (a), with a strong enhancement of the signal below $T_{\mathrm{{c}}}$ at $4\,$K, and an indication that the signal persists into the normal state in agreement with the observations of Friemel \emph{et al.}\cite{friemel_reciprocal-space_2012}, see inset of (a). They interpret this result as indicating a d-wave symmetry for the superconducting pairing state based on calculation by Maier \emph{et al.}\cite{maier_d-wave_2011}. We see no evidence of a magnetic signal at the $(0.5\,0)$ position where the resonance is seen in other iron-based superconductors and was also observed by Wang \emph{et al.\cite{wang_pi0_2012}} in a sample of superconducting Rb$_{0.82}$Fe$_{1.68}$Se$_{2}$.

Finally, we make some remarks about the absolute intensities of the magnetic features. The scattering intensities in our measurements and those of Ref.~\onlinecite{wang_pi0_2012} are calibrated and given in absolute units of cross section. This allows us to compare the strengths of the magnetic signal from the sample of Rb$_{x}$Fe$_{2-y}$Se$_{2}$ used in Ref.~\onlinecite{wang_pi0_2012} with those from the sample of Cs$_{x}$Fe$_{2-y}$Se$_{2}$ used here. The amplitude of the spin-wave peak at 14\,meV for Cs$_{x}$Fe$_{2-y}$Se$_{2}$ (Fig.~\ref{fig:AFM_peaks} above) is about 55 mb\,sr$^{-1}$\,meV$^{-1}$\,f.u.$^{-1}$, which is similar to the amplitude of 40 mb\,sr$^{-1}$\,meV$^{-1}$\,f.u.$^{-1}$ at 10\,meV for Rb$_{x}$Fe$_{2-y}$Se$_{2}$ (Fig.~5(b) of Ref.~\onlinecite{wang_pi0_2012}). However, the amplitude of the resonance peak in Cs$_{x}$Fe$_{2-y}$Se$_{2}$, about 2.5 mb\,sr$^{-1}$\,meV$^{-1}$\,f.u.$^{-1}$, is about five times larger than that reported for Rb$_{x}$Fe$_{2-y}$Se$_{2}$ --- compare Fig.~\ref{fig:ResonancePlot}(b) above with Fig.~6 of Ref.~\onlinecite{wang_pi0_2012}. One should of course be cautious when comparing peak amplitudes. Nevertheless, it does appear that the resonance peak is more prominent in Cs$_{x}$Fe$_{2-y}$Se$_{2}$ than in Rb$_{x}$Fe$_{2-y}$Se$_{2}$. This could indicate that the crystal used here has a higher volume fraction of superconducting phase than that used in Ref.~\onlinecite{wang_pi0_2012}. 

It is also interesting to compare the strength of the resonance peak with that in other Fe-based superconductors. Results for $\chi''(\omega)$ have been reported previously for BaFe$_{1.87}$Co$_{0.13}$As$_2$ and BaFe$_{1.9}$Ni$_{0.1}$As$_2$.\cite{lester_dispersive_2010,liu_nature_2012} In both cases the resonance peak amplitudes (i.e. the increase on cooling below $T_{\rm c}$) are 3--4\,${\mu_{\rm B}}^2$\,eV$^{-1}$\,f.u.$^{-1}$ and the energy-integrated signal $\sim 0.015$\,${\mu_{\rm B}}^2$\,f.u.$^{-1}$. From the inset to Fig.~\ref{fig:ResonancePlot}(a) the corresponding values for Cs$_{x}$Fe$_{2-y}$Se$_{2}$ are ($3.0\pm 0.5$)\,${\mu_{\rm B}}^2$\,eV$^{-1}$\,f.u.$^{-1}$ and ($0.015 \pm 0.003$)\,${\mu_{\rm B}}^2$\,f.u.$^{-1}$, remarkably similar to the values for the two arsenide superconductors. Since the latter were near optimal doping they are expected to be bulk superconductors with close to 100\% superconducting volume fraction. It is tempting, therefore, to conclude that the resonance peak and hence superconductivity in Cs$_{x}$Fe$_{2-y}$Se$_{2}$ is associated with most or all of the sample volume. However, there are many other factors that could control the size of the resonance peak, e.g. the degree of nesting, strength of magnetic correlations, etc. and these may differ from one material to another. We simply note that the resonance peak in Cs$_{x}$Fe$_{2-y}$Se$_{2}$ is similar in strength to that in other Fe-based superconductors.

\section{Conclusions}

The magnetic spectrum of Cs$_{0.8}$Fe$_{1.9}$Se$_{2}$ studied in this work comprises two components: a low energy resonance-like excitation with wave vector $(0.25, 0.5)$ which responds to superconductivity and is similar in strength to the corresponding feature found in other Fe-based superconductors, and spin-wave excitations of the block antiferromagnetic order with wave vector ${\bf Q}_{\rm AFM} = (0.1, 0.3)$ which do not respond to superconductivity to within the experimental sensitivity. The spin-wave component closely resembles that of non-superconducting (insulating) Rb$_{x}$Fe$_{2-y}$Se$_{2}$. Together with other recent studies, these results are consistent with a microstructure composed of spatially separate superconducting and non-superconducting domains, with the $\sqrt{5} \times \sqrt{5}$ Fe vacancy superstructure and block antiferromagnetism confined to the non-superconducting phase. It remains a materials challenge to try to maximize the volume fraction of the superconducting phase.

%In conclusion we have observed the magnetic dynamics in two independent phases of superconducting Cs$_{0.8}$Fe$_{1.9}$Se$_{2}$. The spin dynamics of the $\sqrt{{5}}\times\sqrt{{5}}$ magnetic phase appear to be qualitatively the same as is found in insulating Rb$_{0.89}$Fe$_{1.58}$Se$_{2}$. There is no evidence for a change in the intensity of these spin waves on crossing the superconducting transition temperature. We observe a magnetic resonance at $(0.25\,0.5)$ associated with the superconducting phase of the sample, suggesting that this feature is common throughout the $A_{x}$Fe$_{2-y}$Se$_{2}$ series.

% Acknowledgements
This work was supported by the U.K. Engineering \& Physical Sciences Research Council and the Science \& Technology Facilities Council. Work in Switzerland was supported by the Swiss National Science Foundation and its NCCR programme MaNEP. We thank M. Kenzelmann, A. Podlesnyak, L.-P. Regnault and F. Bourdarot for help with the experimental work.

\bibliographystyle{apsrev4-1}
\bibliography{AxFe2Se2}

\end{document}